\documentclass[journal=jacsat,manuscript=article,email=false,layout=twocolumn]{achemso}

\mciteErrorOnUnknownfalse
\let\oldmaketitle\maketitle         
\let\maketitle\relax                
\usepackage[backgroundcolor= white, linecolor=black, textsize=small]{todonotes}       
\usepackage[fontsize=9pt]{scrextend} 
\usepackage{amsmath}
\usepackage{amssymb}

\author{Adyant Agrawal}
\affiliation[ICP]{Institute for Computational Physics, University of Stuttgart, Stuttgart, Germany}
\author{Catherine Kamal}
\affiliation[UCL]{Department of Mathematics, University College London, London, United Kingdom}
\email{c.kamal@ucl.ac.uk}
\author{Simon Gravelle}
\affiliation[UGA]{Université Grenoble Alpes, CNRS, Laboratoire Interdisciplinaire de Physique (LIPhy), Grenoble, France}
\author{Lorenzo Botto}
\affiliation{Process and Energy Department, Faculty of Mechanical Engineering (3mE), Delft University of Technology, Delft, The Netherlands}
\email{l.botto@tudelft.nl}

\title[Negative intrinsic viscosity]{Negative intrinsic viscosity in graphene nanoparticle suspensions induced by hydrodynamic slip}

\abbreviations{2D, MD, BI}

\begin{document}







\twocolumn[                 
\begin{@twocolumnfalse}     
\oldmaketitle               
\vspace*{-2em}
\begin{abstract}
\noindent The viscosity of nanoparticle suspensions is always expected to increase with particle concentration. However, a growing body of experiments on suspensions of atomically thin nanomaterials such as graphene contradicts this expectation. Some experiments indicate effective suspension viscosity values that fall below that of pure solvent at high shear rates and low solid concentrations, i.e., the intrinsic viscosity is negative. To explain this puzzling phenomenon, we combined molecular dynamics and boundary integral simulations to investigate the shear viscosity of few-nanometer graphene sheets in water at high Péclet numbers (Pe $> 100$). Our results, covering geometric aspect ratios from 4.5 to 12.0, show robustly that the intrinsic viscosity decreases with increasing aspect ratio and becomes negative beyond a threshold aspect ratio $\approx 5.5$. We demonstrate that this anomalous behavior originates from hydrodynamic slip at the liquid-solid interface, which suppresses particle rotation and promotes stable alignment with the flow direction, thereby reducing viscous dissipation relative to dissipation in pure solvent. This slip mechanism holds for both fully 3D disc-like and quasi-2D particle geometries explored in the molecular simulations. As the concentration of graphene particles increases in the dilute regime, the viscosity initially decreases, falling below that of pure water. At higher concentrations, however, particle aggregation becomes significant, leading to a rise in viscosity after a minimum is reached. These findings confirm the occurrence of a negative intrinsic viscosity in a graphene suspension due only to hydrodynamic effects.  Our work has important implications for the design of lubricants, inks, and nanocomposites with tunable viscosity.
\end{abstract}
\vspace*{2em}
\end{@twocolumnfalse}   
]                       

\section{Introduction}
Two-dimensional (2D) materials such as graphene have attracted significant scientific interest over the past two decades due to their exceptional electronic, thermal, mechanical, and optical properties, resulting in new research directions and enabling the development of technologies that are now entering commercial applications.\cite{ferrari2015science,novoselov20162d,zhao2024electrochemical,song2023solution,choi2022large,tyagi2020recent,zeng2018exploring}
Many applications, ranging from lubricants and nanocomposites to conductive inks and coatings, involve the dispersion of 2D material nanoparticles in liquids. Predicting the flow behavior of such suspensions requires quantifying their rheological properties.\cite{kostarelos2014graphene,xiao20172d} Among these, the effective steady shear viscosity is the most important rheological property.

Recent experiments on suspensions of sheet-like materials of nanometric thickness in both simple and complex fluids have revealed rheological behaviors that cannot be explained by classical suspension hydrodynamics\cite{xiao20172d}.
There is increasing evidence that the addition of graphene nanosheets to polymer nanocomposites can reduce the shear viscosity.\cite{kotsilkova2023exploring,DelGiudice2017,White2015}
Relative viscosities (defined as the ratio of effective viscosity to suspending fluid viscosity) well below unity have been reported for suspensions of graphene nanoparticles in lubricating oils in the regime of low particle concentrations.\cite{Bakak2021,pakharukov2022}
A similar trend has been observed for few-layer graphene dispersed in the ionic liquid [HMIM]BF$_4$.\cite{Wang2012b,Liu2014a}
Plate-like nanoparticles of $\alpha$-zirconium phosphate and yttrium oxide, with geometric aspect ratios around 10, have also been reported to reduce the shear viscosity of mineral oil in the dilute regime, with viscosity decreasing as the particle concentration increases.\cite{He2014,He2014a}

These findings are unexpected. Classical suspension rheology predicts that shear viscosity always increases with solid concentration, irrespective of particle shape or Brownian effects.\cite{guazzelli2011physical} Physically, this increase is due to the fact that suspended particles disturb the flow because they impose a boundary condition on the velocity field.  The flow gradients associated with this disturbance translate to an enhanced viscous dissipation compared to the pure solvent. This classical result has been confirmed by extensive theoretical, experimental, and simulation studies since at least the pioneering work of Einstein, who calculated an expression for the viscosity of dilute suspensions of rigid spheres\cite{happel2012}. In contrast,  the molecular dynamics (MD) simulations of this paper unambigously demonstrate that graphene nanoparticles, due to their combination of pronounced hydrodynamic slip and atomic-scale thickness, display a suspension viscosity lower than that of the pure fluid. To our knowledge, this phenomenon is unique to ultra-thin, large-slip particles and has not been observed in conventional colloidal systems.

The anomalous viscosity reduction we report is not only of fundamental interest but also holds promise for technological applications. The development of rheological modifiers that can reduce viscosity, or at least mitigate its increase, could revolutionize the field of lubrication. Approximately 23\% of the global energy consumption (over 100 EJ) is attributed to tribological contacts, with 20\% of this energy lost to friction.\cite{holmberg2017influence} Even modest reductions in the viscosity of nanoparticle-based lubricants could, therefore, yield significant energy savings.

The effective shear viscosity $\eta$ of a dilute suspension of small, rigid particles can be expanded in powers of the particle volume fraction $c$ as
\begin{equation}
    \eta = \eta_{0} \left(1 + \alpha\,c + \beta\,c^{2} + \mathcal{O}(c^{3})\right),
    \label{eq:eta_intrinsic}
\end{equation}
where $\eta_0$ is the viscosity of the pure solvent, $\alpha$ is the intrinsic viscosity, and $\beta$ quantifies inter-particle interactions.\cite{stickel2005fluid} For $c \ll 1$, the leading-order term dominates and $\eta \approx \eta_0 (1 + \alpha c)$. A negative intrinsic viscosity ($\alpha < 0$) thus implies that adding particles reduces the viscosity in the dilute regime.
Recent boundary integral simulations on model plate-shaped particles in unbounded two-dimensional shear flow have shown that $\alpha$ can become negative, provided that (i) the particles are sufficiently thin and (ii) they exhibit hydrodynamic slip with a slip length exceeding a threshold value comparable to the particle thickness.\cite{kamal2024,kamal2024flow} While these continuum simulations offer valuable insights, they are based on idealized particle geometries and do not account for molecular-level effects such as water structuring at the liquid-solid interface or adhesive interactions between particles. Moreover, the continuum approach is only an approximation for graphene, whose thickness is comparable to the size of solvent molecules, challenging the validity of continuum hydrodynamics at this scale. Thus, it remains unclear whether negative intrinsic viscosity can arise in realistic molecular models of graphene. For example, previous molecular dynamics simulations of hexabenzocoronene, a disc-like nanographene with a smaller geometric aspect ratio than those considered here, did not observe negative intrinsic viscosity.\cite{Gravelle2021}

In the current work, we systematically investigate the conditions required for observing a negative intrinsic viscosity.
We employ MD simulations to compute the shear viscosity of suspensions of single-layer graphene nanoparticles with varying aspect ratios in water at high P\'eclet numbers.
Non-functionalized, non-oxidized graphene is chosen as a model 2D material due to its atomic-scale thickness ($\approx 0.5$\,nm for a monolayer in water\cite{stankovich2007synthesis}) and its characteristically large hydrodynamic slip length, which typically ranges from $10$ to $100$\,nm in water, small-chain alcohols, and certain ionic liquids.\cite{radha2016molecular,neek2016,kavokine2022fluctuation}.
A substantial slip length at the solid-liquid interface is a precondition for the occurrence of negative intrinsic viscosity, as demonstrated in recent theoretical studies.\cite{kamal2024,kamal2024flow}
Water is selected as the solvent for our MD simulations owing to the availability of reliable force fields and the presence of high-quality MD data for carbon-based materials in aqueous environments\cite{kavokine2022fluctuation, kannam2011,abascal2005general,horn2004}.
Beyond quantifying the intrinsic viscosity, we also investigate the effect of inter-particle interactions on the second-order coefficient $\beta$ in eq.~\ref{eq:eta_intrinsic}. Our analysis reveals that $\beta$ is highly sensitive to particle geometry and highlights significant discrepancies between MD and continuum predictions, particularly at higher concentrations where molecular-scale effects and particle aggregation become important.

\section{Results and Discussion}

\begin{figure}[ht]
\centering
\includegraphics[width=0.45\textwidth]{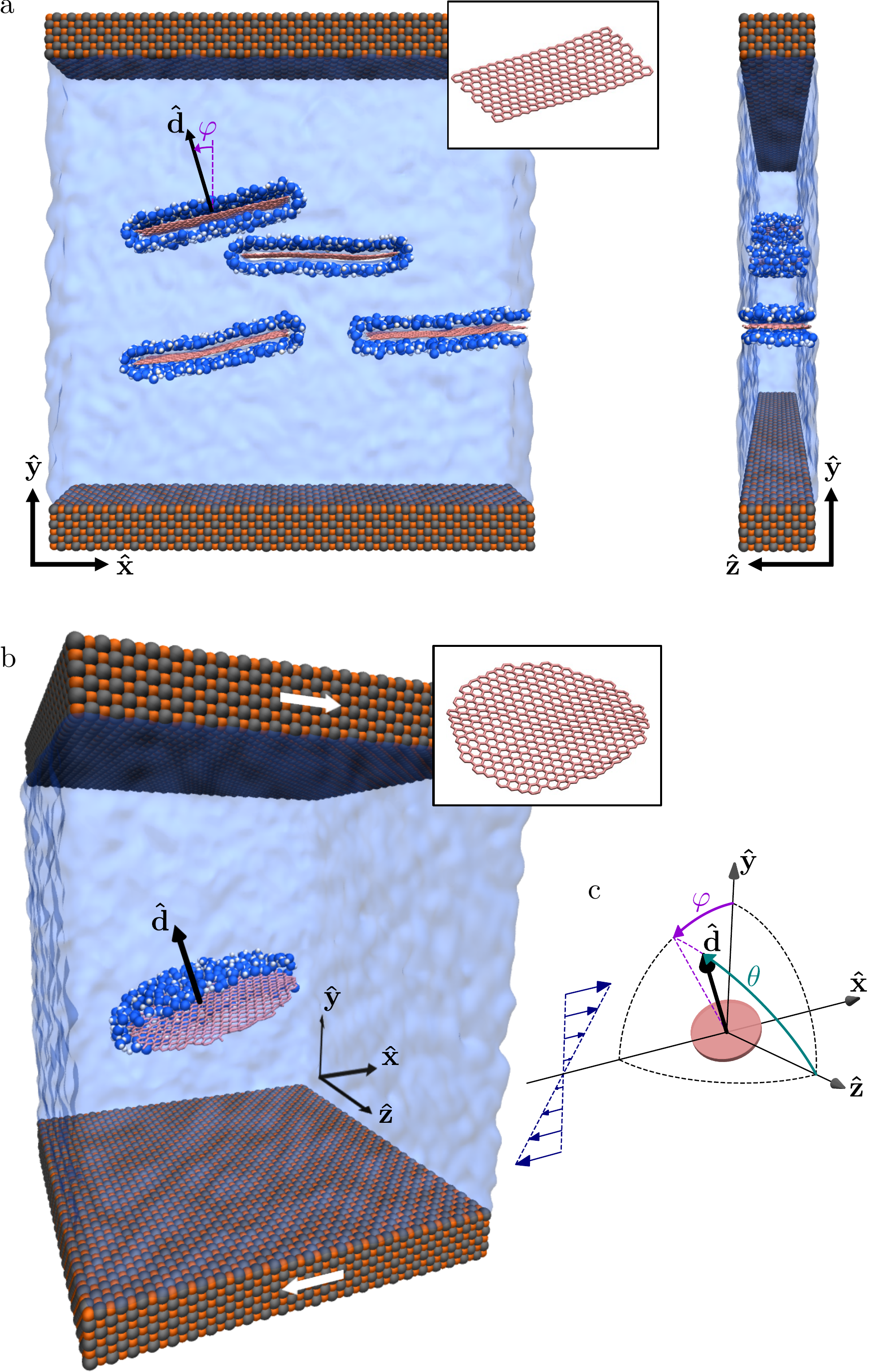}
\caption{
Molecular dynamics simulation setup. The two panels illustrate the two nanoparticle geometries used:
(a) several quasi-2D graphene sheets; 
(b) a fully 3D disc-like graphene sheet. 
Graphene sheets are shown in pink. Selected water molecules near the graphene are rendered explicitly (blue and white spheres for O and H, respectively), with the remaining fluid shown as a translucent isosurface. 
The confining FeO walls (shown in orange/grey) translate along $\mathbf{\hat{x}}$ to apply a simple shear flow. The simulations are periodic along $\mathbf{\hat{x}}$ and $\mathbf{\hat{z}}$.
(c) Schematic definition of the particle orientation angles $\theta$ and $\varphi$, where the unit director vector corresponding to the particle's symmetry axes is given by $\mathbf{\hat d} = (-\sin\theta \sin\varphi,\, \sin\theta \cos\varphi,\, \cos\theta)$.
}
\label{fig:MDsnap}
\end{figure}

MD simulations of plate-like nanoparticles in water were carried out using LAMMPS \cite{LAMMPS} (Fig.\,\ref{fig:MDsnap}). 
The particles were suspended in a shear flow $\boldsymbol{u}^{\infty} = \dot{\gamma} y \mathbf{\hat x}$ produced by two translating walls. Here, $y$ denotes the coordinate along the gradient direction (aligned with $\hat{\mathbf{y}}$) relative to the center of the simulation domain, the flow is directed along $\hat{\mathbf{x}}$, and $\dot{\gamma}$ is the applied shear rate. Periodic boundary conditions were applied along the $\mathbf{\hat x}$ and $\mathbf{\hat z}$ directions. To generate the shear flow, a constant translational velocity was imposed on two solid walls. The walls, composed of iron(II) oxide (FeO), were characterized by a high friction coefficient, ensuring a negligible slip length ($\lambda \approx 0$\,nm) in water.
The nanoparticles were modelled as pristine graphene monolayers and described with the OPLS all-atom force field, which includes explicit bond, angle, and dihedral interactions.\cite{jorgensen1996development,doherty2017revisiting}
Previous molecular dynamics simulations estimated the bending rigidity of graphene in water to be on the order of 1~eV.\cite{gravelle2025effect,agrawal2022viscous}
Based on our selected force fields, the hydrodynamic slip length at the graphene-water interface is estimated to be $\lambda \approx 60$\,nm, consistent with previous molecular dynamics studies.\cite{Kamal2021,Kamal2020}
For comparison, no-slip nanoparticles were simulated by artificially increasing the graphene-water interaction energy, as described in the Methods section. This adjustment effectively eliminated hydrodynamic slip at the particle surface, providing a contrasting behavior to that of pristine graphene.
The system was maintained at a constant temperature of $300$\,K using a Nosé-Hoover thermostat and the fluid was maintained at a pressure of 1~atm by applying a vertical force on the top wall. Further simulation details, including equilibration and production run protocols, are provided in the Methods section.

With this simulation setup, we investigated two nanoparticle geometries: (i) fully 3D disc-like graphene nanosheets (fig.~\ref{fig:MDsnap}b), which are free to rotate and translate in all spatial directions, and (ii) quasi-2D rectangular nanosheets that extend across the entire simulation box along the $\mathbf{\hat z}$ direction (fig.~\ref{fig:MDsnap}a), effectively confining their motion and rotation to the $\mathbf{\hat x}$-$\mathbf{\hat y}$ plane due to periodic boundary conditions. In both geometries, the graphene sheets retain flexibility and can undergo deformation in response to the applied shear flow.


The effective suspension viscosity, $\eta$, was determined by measuring the shear stress on the confining walls. Simulations were first performed with a single particle, corresponding to a dilute solid fraction of approximately $c \simeq 0.01$. At this low concentration, the reduced viscosity, $(\eta/\eta_0 - 1)/c$ provides a good approximation to the intrinsic viscosity $\alpha$.
We investigated a range of particle aspect ratios ($4.5<a/b<12$) for a fixed shear rate of $\dot\gamma = 5 \times 10^{10}$\,s$^{-1}$, where $2a$ and $2b$ denote the effective length (or disc diameter) and width of the particle, respectively. We defined these two dimensions taking into account the effective size of the constituent atoms \cite{Gravelle2021}.
The corresponding P\'eclet number, $\text{Pe} = \dot\gamma / D_\mathrm{r}$, ranged from 80 to 2800, with the lowest P\'eclet number corresponding to the shortest particle. The rotational diffusion coefficient, $D_\text{r}$, was estimated for quasi-2D graphene particles using the expression $D_\text{r} \approx k_\text{B} T / (2 \pi \eta a^3)$ developed for an infinitely thin 2D rigid plate\cite{Sherwood1991}. For 3D particles we used $D_\text{r} = 3 k_\text{B} T / (32 \eta a^3)$\cite{Sherwood2012}. Here, $k_\text{B}$ is the Boltzmann constant, $T$ is the temperature, and $a$ is the particle's half-length.
For $\text{Pe} = \mathcal{O}(10^2)$, the rotational dynamics is controlled by hydrodynamics, and the influence of Brownian noise on the orientational distribution function is minimal \cite{Kamal2021}.

\begin{figure}[ht]
    \centering
    \includegraphics[width=0.45\textwidth]{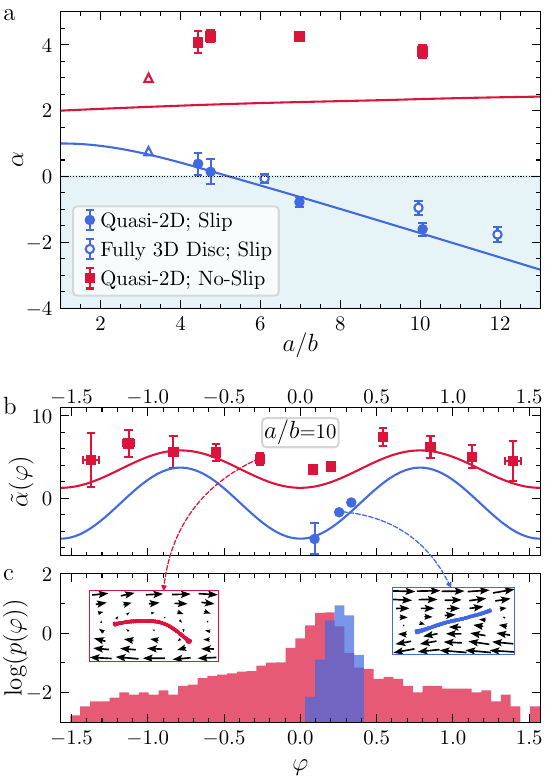}
    \caption{
     Single-particle simulation results for slip (blue) and no-slip (red) graphene particles.
    (a) $\alpha$ vs $a/b$ for fully 3D (open symbols) and quasi-2D (filled symbols) particles. Triangles indicate data from Ref.~\citenum{Gravelle2021}. Lines are 2D BI predictions. Shaded blue represents $\alpha < 0$.
    (b) Conditional intrinsic viscosity $\tilde\alpha(\varphi)$ for quasi-2D particles with $a/b = 10$. The error bars represent the standard error.
    (c) Logarithm of the orientation probability distribution $\log p(\varphi)$ for the same systems. To illustrate how $\tilde\alpha(\varphi)$ reflects the hydrodynamic resistance generated by a particle, in (c) we include insets showing MD snapshots of particle configurations and the surrounding fluid velocity fields.
    }
    \label{fig:alpha}
\end{figure}

Figure~\ref{fig:alpha}a shows the intrinsic viscosity as a function of the particle aspect ratio for both slip and no-slip particles. For slip particles, $\alpha$ decreases monotonically with increasing aspect ratio, becoming negative at $a/b \simeq 5.5$. 
In contrast, no-slip particles consistently exhibit positive values of $\alpha$, with a much weaker dependence on $a/b$ compared to slip particles. Across the range of $a/b \in [4,10]$ considered, $\alpha$ remains approximately constant at $\alpha \approx 4$, with a minor reduction observed at the highest aspect ratio examined ($a/b = 10$).
This reduction coincides with enhanced particle alignment in the flow direction and less frequent tumbling at higher aspect ratios, as evidenced by the molecular dynamics simulations presented in the Supporting Information.

Overall, Fig.~\ref{fig:alpha}a underscores the dramatic reduction in $\alpha$ due to hydrodynamic slip across all aspect ratios, with the effect becoming more pronounced at larger $a/b$. Notably, while previous boundary element simulations by \citet{Allison1999} for ellipsoidal particles at low P\'eclet numbers also reported a significant reduction in $\alpha$ due to slip, the intrinsic viscosity remained positive even for `perfect slip' ellipsoids with $\lambda \to \infty$. For example, for $a/b = 10$, Allison observed a decrease in $\alpha$ from $7.9$ to $1.7$ for the no-slip and perfect slip cases, respectively. In contrast, our results for high P\'eclet numbers reveal a change in sign of the intrinsic viscosity, with $\alpha$ dropping from $3.8 \pm 0.2$ to $-1.6 \pm 0.2$ for quasi-2D particles and from $7.4 \pm 0.5$ to $-1.0 \pm 0.2$ for 3D disks.


To further investigate the mechanism responsible for the negative intrinsic viscosity, we examine the orientation statistics of the particles and the intrinsic viscosity conditioned on the particle orientation.
In simple shear flow, the $xy$-component of the time-averaged macroscopic shear stress, $\langle \Sigma_{xy} \rangle$, comprises two contributions: the solvent stress, $\eta_0 \dot{\gamma}$, and the particle-induced stress, $\langle \Sigma_{p,xy} \rangle$ \cite{guazzelli2011physical}. In the dilute limit, the particle contribution can be expressed as $\langle \Sigma_{p,xy} \rangle = n \langle S_{xy} \rangle$, where $\langle S_{xy} \rangle$ is the $xy$-component of the time-averaged stresslet tensor for an isolated particle, and $n$ is the particle number density \cite{guazzelli2011physical}.
At large Pe, the stresslet is determined primarily by hydrodynamic stresses and represents the deviatoric component of the hydrodynamic traction on the particle's surface \cite{kamal2024}. In Stokes flow, the instantaneous traction at time $t$ depends solely on the particle's configuration at that time. For simplicity, we restrict our attention to a quasi-2D particle whose surface normal ($\hat{\mathbf{d}}$) is constrained to lie within the flow-gradient ($\hat{\mathbf{x}}$-$\hat{\mathbf{y}}$) plane. 
In this configuration, the particle's orientation is fully characterized by the angle $\varphi$, defined as the counterclockwise angle between $\hat{\mathbf{d}}$ and $\hat{\mathbf{y}}$. The time-averaged shear stresslet is expressed as an average of the stresslet $S_{xy}(\varphi)$, conditioned on the particle adopting orientation $\varphi$ and weighted by the corresponding probability distribution $p(\varphi)$:
\begin{equation}
\langle S_{xy} \rangle = \int_{-\pi/2}^{\pi/2} S_{xy}(\varphi)\, p(\varphi)\, \mathrm{d}\varphi.
\end{equation}
The orientational space $\varphi \in [-\pi/2, \pi/2]$ represents the range of possible angles explored by the particle. Similarly, the intrinsic viscosity, $\alpha$, is obtained by averaging the orientation-dependent intrinsic viscosity, $\tilde\alpha(\varphi)$:
\begin{equation}
\alpha = \int_{-\pi/2}^{\pi/2} \tilde\alpha(\varphi) p(\varphi) \, \mathrm{d}\varphi.
\end{equation}
This expression highlights the critical role of particle orientation in determining the shear viscosity of the suspension.

Using MD simulations, we calculated the conditional intrinsic viscosity $\tilde\alpha(\varphi)$ and the orientation probability distribution $p(\varphi)$ for quasi-2D particles. The results, shown in Fig.~\ref{fig:alpha}b,c, correspond to particles with $a/b = 10$, for $\dot{\gamma} = 5 \times 10^{10} \, \mathrm{s}^{-1}$.
The conditional intrinsic viscosity was determined by measuring the wall shear stress at instants when the particle adopted a given orientation angle $\varphi$, while the orientational distribution $p(\varphi)$ was computed by histogramming the frequency of $\varphi$ over the simulation trajectory.

No-slip particles exhibit an approximately cosinusoidal variation in $\tilde\alpha(\varphi)$, with a minimum at $\varphi = 0$ (Fig.~\ref{fig:alpha}b). Unlike no-slip particles, which sample the full range of orientations due to continuous tumbling, slip particles perform fluctuations near a fixed angle, resulting in a peaked profile of $p(\varphi)$ (Fig.~\ref{fig:alpha}c). The value of $\varphi$ about which the fluctuations occur is known to be governed by the particle aspect ratio and by the hydrodynamic slip length~\cite{Kamal2020}.
Because $\tilde\alpha(\varphi)<0$ within the narrow range for which $p$ is non-zero, the time-averaged intrinsic viscosity $\alpha$ is also negative.

The insets of Fig.~\ref{fig:alpha}c, which show representative snapshots from the MD simulations, describe typical particle configurations and the corresponding fluid velocity fields. 
These images clearly demonstrate that slip particles induce significantly less disturbance in the surrounding flow compared to no-slip particles.
The deformation of slip graphene particles is small in comparison to that of no-slip graphene. This observation is explained by the higher shear stress for buckling of slip sheets with respect to no-slip sheets~\cite{gravelle2025effect,kamal2021alignment}.

We compared our MD simulations with non-Brownian, single-particle boundary integral (BI) calculations for quasi-2D rigid particles. The details of BI calculations are given in the Methods section. Note that the use of non-Brownian BI calculations is justified by the weak dependence of $\alpha$ on $\text{Pe}$ for $\text{Pe}\gtrsim100$ regardless of the slip length; the effect of $\text{Pe}$ is discussed further in Supporting Information.

The MD results are in excellent agreement with BI predictions in the slip case ($\lambda=60$~nm) across all aspect ratios (full lines in Fig.~\ref{fig:alpha}a). For no-slip particles ($\lambda=0$~nm), MD simulations predict larger values of $\alpha$ than the BI calculations.
This difference could be explained by the large interaction energy between carbon and oxygen atoms used to enforce the no-slip boundary condition in our MD calculations. A high solid-liquid interaction energy results in water molecules being strongly adsorbed to the graphene platelet, effectively `sticking' to it and leading to an increased effective thickness. This can be observed from the radial distribution function of oxygen atoms around carbon atoms presented in the Supporting Information. It is known that strongly structured fluid molecules around a particle result in higher viscosity\cite{sendner2009interfacial}.

We also computed $\tilde\alpha(\varphi)$ using BI simulations by evaluating the instantaneous single-particle stresslet, given by $\tilde\alpha(\varphi) = {S_{xy}(\varphi)}/{(\eta_0 \dot{\gamma} A_p)}$, where $A_p$ is the particle's cross-sectional area (solid lines in Fig.~\ref{fig:alpha}b).
The close quantitative agreement between the MD and BI predictions shows that both atomistic and continuum simulations describe the same orientation-dependent contributions to the viscosity, reinforcing the robustness of our approach.

\begin{figure}
    \centering
    \includegraphics[width=0.45\textwidth]{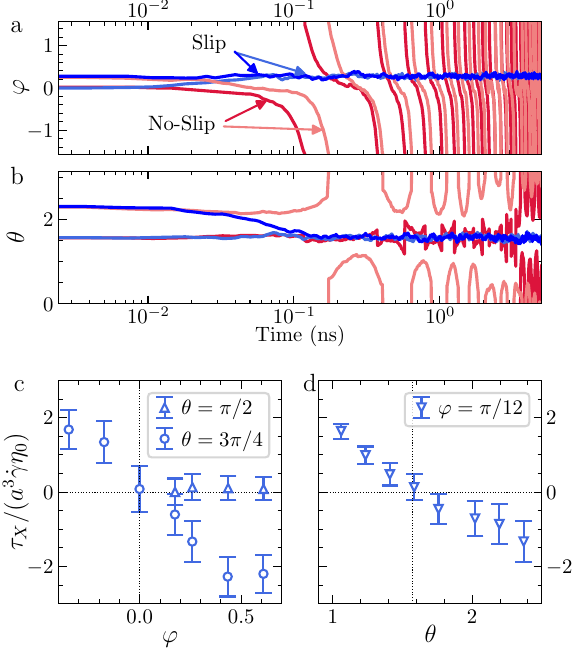}
    \caption{
    Orientation dynamics and hydrodynamic torque for a fully 3D disc-like particle.
    (a, b) Time evolution of (a) azimuthal angle $\varphi$ and (b) polar angle $\theta$ for slip (blue) and no-slip (red) particles, for two initial orientations.
    (c, d) x-component of the hydrodynamic torque for a disc held at fixed orientation, plotted as a function of: (c) $\varphi$, for two values of $\theta$; and (d) $\theta$, for a fixed value of $\varphi$.}
    \label{fig:inclination}
\end{figure}

Our MD simulations reveal minimal difference in shear viscosity between fully 3D and quasi-2D particle geometries. This agreement arises because of the propensity of a fully 3D slip graphene to align on average with its normal in the flow plane.
We describe the instantaneous orientation of each fully 3D graphene particle using the unit vector $\hat{\mathbf{d}}$, which is aligned with the particle's axis of symmetry. For brevity, we refer to the particle's orientation simply as the orientation of $\hat{\mathbf{d}}$.
{To compute $\mathbf{\hat d}$, we apply principal component analysis (PCA) to the centered atomic coordinates of the disc and define $\mathbf{\hat d}$ as the eigenvector corresponding to the smallest eigenvalue of the covariance matrix, following Hoppe et al.\cite{hoppe1992surface}} To uniquely define the director, we choose the sign of ${\mathbf{\hat d}}$ such that ${\mathbf{\hat d}} \cdot {\mathbf{\hat y}} > 0$.
In a spherical coordinate system, the components of ${\mathbf{\hat d}}$ are 
\begin{equation}
    \mathbf{\hat d} = (-\sin\theta \sin\varphi,\, \sin\theta \cos\varphi,\, \cos\theta),
\end{equation}
where $\theta$ is the polar angle with respect to the vorticity direction ${\mathbf{\hat z}}$, and $\varphi$ is the azimuthal angle in the flow-gradient (${\mathbf{\hat x}}$-${\mathbf{\hat y}}$) plane (see Fig.~\ref{fig:MDsnap}c).
Our MD simulations show that, regardless of their initial orientation, slip discs rapidly (within the first few hundred picoseconds) relax to an orientation for which $\theta \simeq \pi/2$ and $\varphi \ll 1$ (Fig.~\ref{fig:inclination}a,b). This stable alignment of particle with the flow-gradient plane is consistent with previous MD simulations of nanographene reported by \citet{Gravelle2021}. The observation that a single 3D graphene disk eventually aligns such that its symmetry axis lies in the flow-gradient plane, performing a motion that is statistically two-dimensional, helps understanding why the quasi-2D particle simulations (Fig.\ref{fig:MDsnap}a ) and the fully 3D disk simulations (Fig.\ref{fig:MDsnap}b ) give quantitatively comparable results.

In contrast, the no-slip discs exhibit persistent tumbling and periodic oscillations in both $\varphi$ and $\theta$, consistent with Jeffery's classical theory (Fig.~\ref{fig:inclination}a-b). Occasionally, these particles transition between Jeffery orbits, gradually shifting from tumbling towards orbits closer to $\theta = \pm \pi/2$. This transition is facilitated by weak Brownian rotational diffusion, which enables the particle to explore a range of Jeffery orbits. Such behavior is well-described by the theoretical framework of Leal and Hinch~\cite{Leal1971}, which extends Jeffery's theory to account for the influence of weak Brownian noise.

To further understand the tendency of graphene particle to reorient towards a stable orientation, we computed the hydrodynamic torque component along the flow direction, $\tau_X$, exerted by the fluid on a particle held at fixed orientation (shear rate $\dot{\gamma} = 5 \times 10^{10} \, \mathrm{s}^{-1}$).
Figure~\ref{fig:inclination}c compares $\tau_X$ as a function of the azimuthal angle $\varphi$ for two polar angles. 
For $\theta = \pi/2$, the torque remains nearly zero for the entire range of $\varphi$, consistent with a stable in-plane orientation. In contrast, for a more tilted configuration ($\theta = 3\pi/4$), the torque exhibits a pronounced dependence on $\varphi$. Specifically, $\tau_X$ is negative for $\varphi > 0$, indicating a restoring torque towards the flow-gradient plane. While $\tau_X$ is positive for $\varphi < 0$, indicating a torque that drives the particle further from the flow-gradient plane.
Fig.~\ref{fig:inclination}d illustrates the dependence of $\tau_X$ on $\theta$ for a fixed azimuthal angle $\varphi = \pi/12$, which corresponds to the steady-state orientation typically observed for slip discs in our simulations (see Fig.~\ref{fig:inclination}a). The torque is positive for $\theta < \pi/2$, negative for $\theta > \pi/2$, and vanishes at $\theta = \pi/2$, indicating a restoring hydrodynamic moment that drives the particle back toward $\theta = \pi/2$ when tilted out of the flow-gradient plane.


\begin{figure}
    \centering
    \includegraphics[width=0.48\textwidth]{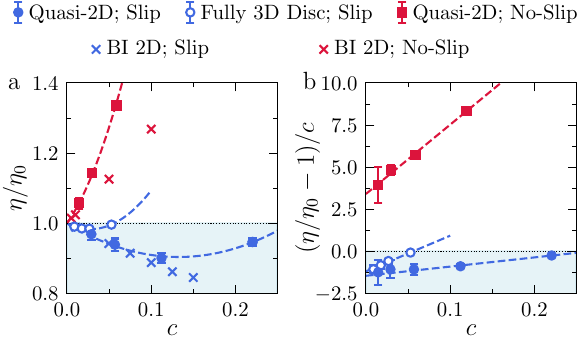}
    \caption{(a) Relative shear viscosity vs solid fraction for fully-3D and quasi-2D particles with $a/b = 10$. The crosses represent BI simulation results~\cite{kamal2024flow}. (b) Reduced shear viscosity values corresponding to the data in (a). Dashed lines are least-squares fits to the data.}
    \label{fig:beta}
    \end{figure}

\begin{table}
    \centering
    \begin{tabular}{lccc}
    \hline
    \textbf{Particle Type} & \(\alpha\) & \(\beta\) & \(c_{\text{min}} = -\alpha/(2\beta)\) \\
    \hline
    Quasi-2D No-Slip  & \(3.4\)   & \(41.1\)  & N/A \\
    Quasi-2D Slip     & \(-1.4\)  & \(5.4\)   & \(0.13\) \\
    Fully-3D Slip     & \(-1.2\)  & \(21.1\) & \(0.03\) \\
    \hline
    \end{tabular}
    \caption{Coefficients $\alpha$, $\beta$ and minimum viscosity concentration \(c_{\text{min}}\) (where applicable) for graphene particles with $a/b=10$.}
\label{tab:viscosity_params}
\end{table}

To explore the influence of inter-particle interactions on the suspension viscosity, we simulated suspensions with solid fractions in the range $c \in [0.02, 0.22]$ by varying the number of particles while keeping the wall separation constant.
For both fully-3D and quasi-2D particles, the effective dimensions were $2a \approx 5$\,nm (length) and $2b \approx 0.5$\,nm (width).
Fig.~\ref{fig:beta}a shows the relative viscosity $\eta/\eta_0$ as a function of the solid fraction $c$.
The simulations were performed at a shear rate of $\dot\gamma = 7 \times 10^9$\,s$^{-1}$, corresponding to a P\'eclet number on the order of $10^2$.
Our findings indicate that, even in the semi-dilute concentration regime, suspensions of slip particles exhibit a relative viscosity below unity (blue symbols in Fig.~\ref{fig:beta}a).
However, the dependence of $\eta/\eta_0$ on the solid fraction $c$ becomes non-monotonic: the viscosity initially decreases from 1, in agreement with the dilute-limit prediction, but then increases again beyond a critical concentration.

In Fig.~\ref{fig:beta}b, we plot the reduced viscosity $\left(\eta/\eta_0-1\right)/c$ as a function of $c$.
For both slip and no-slip particles, the reduced viscosity was found to increase linearly with $c$. We calculated the linear and quadratic coefficients $\alpha$ and $\beta$ in eq.~\ref{eq:eta_intrinsic} by fitting the MD data for the reduced viscosity to a line (dashed lines in Fig.~\ref{fig:beta}a,b).
The obtained values of $\alpha$ and $\beta$ are listed in Table~\ref{tab:viscosity_params}.

For quasi-2D particles, $\beta$ is significantly larger for no-slip particles compared to slip particles. This difference stems from different particle-particle interaction behaviours: no-slip particles exhibit collective orbital motion, leading to frequent collisions, whereas slip particles primarily slide past one another with minimal interactions (see Supporting Information).
Interestingly, the values of $\alpha$ for quasi-2D and fully-3D particles are quantitatively similar, but this is not the case for values of $\beta$. Fully-3D slip discs exhibit significantly larger $\beta$ values compared to quasi-2D slip particles.
This difference results from the enhanced tendency of fully-3D discs to aggregate, because of their increased rotational freedom and stronger edge-mediated interactions.

In addition to single-particle BI calculations, we compare our MD data with multi-particle BI data presented in Kamal and Botto~\cite{kamal2024flow}(crosses in Fig.~\ref{fig:beta}a). Both BI and MD show that the reduced viscosity increases with solid fraction at a slower rate for slip particles compared to no-slip particles. However, the multi-particle BI calculations predict smaller values of the inter-particle interaction term, $\beta$, and do not capture the minimum in effective viscosity observed in the MD simulations at intermediate concentrations.
These differences are due to the numerical treatment of near-contact particle-particle interactions in the BI model, where a short-range repulsive force was used to avoid overlap between the particles.
Consequently, the results of the multi-particle BI calculations are most applicable to graphene nanosheets dispersed in good solvents (or in the presence of small-molecule dispersants), where adhesive interactions between the sheet-like nanoparticles are minimal.

The negative value of $\alpha$ for slip particles leads to a minimum in the suspension viscosity at $c_{\text{min}} \approx -\alpha/(2\beta)$. This concentration is smaller for fully-3D particles, consistent with their larger $\beta$ values.
These findings suggest that, for achieving substantial reductions in relative viscosity at finite solid concentrations using slip particles, it is essential to minimize $\beta$ by suppressing adhesive inter-particle interactions.

\section{Conclusion}

We have systematically investigated the shear viscosity of suspensions of nanometer-scale graphene sheets using MD simulations, complemented by continuum BI simulations of the Stokes equations. Our results demonstrate that, for sufficiently large P\'eclet numbers, graphene particles (sheets) with aspect ratios above approximately 5.5 reduce the effective suspension viscosity below that of the pure solvent. This provides the first direct, molecular-level confirmation of prior continuum predictions and supports recent experimental observations of anomalous viscosity reduction in graphene-based suspensions.
In the dilute regime, where particle interactions are negligible, MD and BI simulations are in excellent agreement, confirming that the phenomenon is driven by hydrodynamic slip and particle alignment rather than simulation artifacts or experimental uncertainties.
The absence of negative intrinsic viscosity in a prior MD study~\cite{Gravelle2021} is attributed to the use of nanoparticles with insufficient aspect ratio.
At higher concentrations, MD simulations yield a larger interaction coefficient ($\beta$) than BI simulations, primarily due to the aggregation tendency of graphene nanoparticles in water, a feature not captured in the continuum BI model.

In dilute suspensions, particles with both quasi-2D and fully-3D geometries yield comparable intrinsic viscosity values, validating the quasi-2D approximation commonly used in previous studies~\cite{Kamal2020,Kamal2021,kamal2021alignment,Gravelle2021}.
In both geometries, slip particles stabilize with their symmetry axis aligned near the gradient direction, whereas no-slip particles undergo continuous tumbling as described by Jeffery's theory.
The negative intrinsic viscosity arises from the interplay between hydrodynamic slip and stable particle alignment. Thin, well-aligned slip particles minimally disturb the surrounding fluid, and the velocity gradients along their slender surfaces are substantially reduced compared to those in undisturbed flow.
This mechanism is analogous to that in high-capillary-number, low-Reynolds-number bubbles, which also exhibit negative intrinsic viscosity~\cite{rust2002effects}, though bubbles have effectively infinite slip length, unlike graphene.

For particles with slip, the interaction coefficient $\beta$ is reduced by approximately an order of magnitude compared to no-slip particles. This reduction arises because the stable alignment of slip particles allows them to slide past one another while maintaining nearly parallel orientations. As a result, inter-particle interactions remain weak even when the center-to-center separation is comparable to the particle length.
In our simulations, particle-particle interactions are more pronounced for fully-3D disc-like particles than for their quasi-2D counterparts due to the increased rotational freedom of the 3D particles.
Previous theoretical studies have reported the influence of slip length on the hydrodynamic interaction coefficient $\beta$ in idealized systems. For example, \citeauthor{luo2007interception}~\cite{luo2007interception} computed $\beta$ for pairwise interactions of slip spheres, while \citeauthor{kamal2024flow}~\cite{kamal2024flow} considered quasi-2D circular cylinders with infinite depth.
In both studies, slip has been found to reduce the effective interaction strength; however, those analyses neglect aggregation and anisotropic orientation effects, which are instead present in the current work.

The close agreement between MD and BI simulations indicates that our dilute-limit findings are broadly applicable to other solvents and 2D materials beyond graphene, provided that the key continuum parameters---the slip length to thickness ratio ($\lambda/b$), the aspect ratio ($a/b$) and the P\'eclet number (Pe)---are sufficiently large (see Table 1 in Ref.~\citenum{Kamal2021} for a survey of slip lengths across various systems).
However, in the semi-dilute and dense regimes, the surface properties of the particles---and thus their tendency to adhere to each other, an effect which is particularly important for sheet-like particles---become significant.

A key challenge in realizing negative intrinsic viscosity at technologically relevant concentrations lies in the strong tendency of graphene to aggregate, especially in water. Future work should therefore investigate how molecular dispersants or alternative solvents influence both the dispersion stability and the suspension viscosity of 2D materials. Another important factor is flexibility: while the nanometer-scale graphene flakes studied here are relatively rigid, larger micrometer-scale sheets may experience significant bending, wrinkling, or self-folding under flow~\cite{annett2016,salussolia2022simulation,funkenbusch2024dynamics,yu2022wrinkling,silmore2021buckling}, potentially leading to a trend towards a positive intrinsic viscosity.
Since slip reduces compressional viscous forces, the buckling transition becomes less significant for slip sheets than for no-slip sheets under practical shear rates~\cite{kamal2021alignment}.
These considerations highlight the complex interplay between flexibility, slip, and hydrodynamics in large graphene sheets—a topic that remains to be fully understood. Our results offer a framework to enhance understanding of the hydrodynamic behavior of atomically thin materials.

\section{Methods}

\subsubsection{Molecular dynamics simulation details}

The TIP4P/2005 model was used for water \cite{abascal2005general}, and the OPLS-AA force field was used for the graphene nanosheets \cite{doherty2017revisiting,jorgensen1996development}.
By default, carbon-water, water-wall and carbon-wall interaction parameters were calculated using the Lorentz-Berthelot mixing rules.
In that case, the hydrodynamics slip length at the graphene-water interface was determined to be $\lambda = 60 \pm 11$~nm following the method described in \citet{herrero2019shear}.
To model no-slip boundary conditions at the graphene-water interface, the Lennard-Jones interaction energy between carbon atoms in graphene and oxygen atoms in water was adjusted to suppress interfacial slip, yielding a slip length of $\lambda \lesssim 0.1$~nm (see Supporting Information for details).
An instantaneous vertical force was applied to the top wall, while the bottom wall was constrained in the vertical direction, ensuring a pressure of 1~atm in the solution. A Nos\'e-Hoover temperature thermostat \cite{nose1984molecular,hoover1985canonical} was applied to the degree of freedom normal to the direction of the flow to maintain a constant temperature of $T=300$\,K in the solution.
The system was equilibrated without moving the walls for 200 ps. 
The walls were then moved at a constant velocity along the $\hat{x}$ direction for a second equilibration phase of 100~ps, after which production runs were conducted. 
During these runs, the force exerted on the walls, along with the positions and velocities of all atoms in the system, were recorded.

\subsubsection{Continuum boundary integral formulation}

We employ a boundary integral method to solve the incompressible Stokes equations for a non-Brownian, isolated, rigid, quasi-2D particle freely suspended in the imposed flow field $\boldsymbol{u}^{\infty}$. Since the quasi-2D particle is effectively two-dimensional, the method requires the specification of a continuous line boundary $\mathcal{L}$ onto which the integral equations are discretized. Molecular dynamics simulations of the flow field around a single-layer quasi-2D graphene in water suggest that $\mathcal{L}$ can be well-approximated as a rectangle with semi-circular edges \cite[]{Kamal2020,Kamal2021}.
The Navier slip boundary condition is applied on $\mathcal{L}$, where the slip velocity is given by
\begin{equation}
    \boldsymbol{u}^{\text{sl}} = \frac{\lambda}{\eta} \boldsymbol{n} \times \boldsymbol{f} \times \boldsymbol{n}, \label{eq:slip}
\end{equation}
where $\boldsymbol{n}$ is the outward-pointing normal vector, $\boldsymbol{f}$ is the hydrodynamic traction distribution over $\mathcal{L}$, $\lambda$ is the slip length, and $\eta$ is the dynamic viscosity of the suspending fluid.

In the two-dimensional boundary integral method, the incompressible Stokes equation is recast as an integral over $\mathcal{L}$. For a point $\boldsymbol{x}_1\in \mathcal{L}$, the boundary integral equation is \cite{pozrikidis1992boundary,kamal2024}
\begin{equation}
\begin{aligned}
    &\frac{1}{4\pi}\int_\mathcal{L} \boldsymbol{n}(\boldsymbol{x})\cdot \boldsymbol{K}(\boldsymbol{x}-\boldsymbol{x}_1)\cdot \boldsymbol{u}^{\text{sl}}(\boldsymbol{x}) \text{d}L(\boldsymbol{x})\\
    &-\frac{1}{4\pi\eta}\int_\mathcal{L}\boldsymbol{G}(\boldsymbol{x}-\boldsymbol{x}_1)\cdot\boldsymbol{f}(\boldsymbol{x})\text{d}L(\boldsymbol{x})\\
    &=\frac{\boldsymbol{u}^{\text{sl}}(\boldsymbol{x}_1)}{2}+\Omega \mathbf{\hat y}\times \boldsymbol{x_1}-\boldsymbol{u}^{\infty}(\boldsymbol{x}_1), 
\end{aligned}\label{eq_bieq}
\end{equation}
where $\boldsymbol{G}$ and $\boldsymbol{K}$ are tensors associated with the two-dimensional Stokeslet and stresslet, respectively \cite{pozrikidis1992boundary} and $\text{d}L(\boldsymbol{x})$ is a boundary element. We have assumed that the centre of the particle is positioned at the origin, so that the particle rotates with angular velocity $\Omega \mathbf{\hat y}$. 

Equation~\eqref{eq_bieq} is solved numerically, along with the condition of zero net torque, to find $\boldsymbol{f}$ and $\Omega$ for a specific orientation angle $\varphi$. The numerical scheme is described and validated in Ref.\citenum{kamal2024}. 

In what follows, we will show that the intrinsic viscosity coefficient $\alpha$ can be evaluated in terms of $\boldsymbol{f}$. For a system of two-dimensional particles, $\alpha$ can be expressed as
\begin{equation}
   \alpha=A\left<1-\cos{4\varphi}\right>+B \label{eq:stress_h}, 
\end{equation}
where $A,\;B$ are dimensionless coefficients and the angled brackets $\left<\;\right>$ represent an average over the steady-state orientation distribution function $p(\varphi)$.

The coefficients $A,~B$ depend on the particle shape and slip length. Using matrix transformations shown in Refs. \citenum{Kamal2021} and \citenum{kamal2024}, the coefficients $A$ and $B$ can be decomposed as
\begin{equation}
\begin{split}
A &=\frac{S_{ss}(\pi/4) - S_{tt}(\pi/4)
- 2S_{st}(0)} {4\dot{\gamma}\eta {A}_p},\\
B &=\frac{S_{st}(0)}{\dot{\gamma}\eta {A}_p},
\label{eq:ABstress}
\end{split}
\end{equation}
where $A_p$ is the cross-sectional area of the particle. 
These coefficients are expressed in terms of the hydrodynamic stresslet tensor $S_{ij}$. This tensor is evaluated in terms of $\boldsymbol{f}$ as \cite{pozrikidis1992boundary} 
\begin{equation}
\begin{split}
S_{ij}(\varphi)=\frac{1}{2}\int_{{\mathcal{L}}} \left[ f_i(\varphi) x_j+f_j(\varphi) x_i\right. \\\left. -2\eta(u^{\text{sl}}_i n_j + u^{\text{sl}}_j n_i)\right] \mathrm d{L},
\end{split}
\end{equation}
 In this equation, $S_{ij}$ is evaluated in the particle frame ($\mathbf{\hat s},\mathbf{\hat t}$), where the unit vectors $\mathbf{\hat s}$ and $\mathbf{\hat t}$ are parallel to the platelet's major and minor axes, respectively.

For Pe$\to\infty$, $\left<1-\cos{4\varphi}\right>$ in eq.~\eqref{eq:stress_h} can be evaluated analytically in terms of $k_e$ as \cite[]{Kamal2021}:
\begin{equation}
= \left\{
\begin{array}{ll}
	4k_\text{e}\left(k_\text{e}+1\right)^{-2}, &\text{if }k_\text{e} \in \mathbb{R}, \\ 1-\cos{\left(4\arctan{|k_e|}\right)}, &\text{if } k_\text{e} \in i\mathbb{R}.
\end{array}
\right.
\label{eq:cos4t}
\end{equation} 
Here, $k_e=\sqrt{T(0)/T(\pi/2)}$ is the square root of the ratio between the torques exerted on a particle held fixed parallel ($T(0)$) and perpendicular ($T(\pi/2)$) to the flow. For no-slip quasi-2D particles, $k_e$ is always real \cite{Kamal2021} and follows a power-law relationship with the geometric aspect ratio \cite[]{singh2014rotational}. Increasing $\lambda$, reduces $k_e$ so that $k_e=0$ at a critical slip length $\lambda_c\sim b$ \cite[]{Kamal2020,Kamal2021}. The cause for this reduction is that slip reduces the tangential traction over the slender region of the particle surface when the particle is held fixed in the direction of flow, which decreases the value of $T(0)$. As $\lambda/a\to \infty$, the contribution to $T(0)$ originating from the slender portion of the particle vanishes, and the resulting torque comes from the contribution from the edges. The contribution from the edges has an opposite sign to $T(\pi/2)$, resulting in $k_e\propto i\sqrt{k}$ in this limit \cite{Kamal2020}.


\begin{acknowledgement}

L.B. acknowledges funding from the European Research Council (ERC) under the European Union's Horizon 2020 research and innovation program (grant agreement N$^\circ\;715475$, project FlexNanoFlow). S.G. acknowledges funding from the European Union's Horizon 2020 research and innovation programme under the Marie Skłodowska-Curie grant agreement N$^\circ\;101065060$, project NanoSep. C.K. acknowledges funding from the Royal Society Dorothy Hodgkin fellowship, grant number G119482.

\end{acknowledgement}


\begin{suppinfo}BI calculations of intrinsic viscosity at finite Péclet numbers and associated analysis; method for determining hydrodynamic slip length in MD simulations and tuning of slip length via graphene–water interaction strength; hydrodynamic flow fields around particles; and additional MD results on particle orientation statistics (PDF).
\end{suppinfo}

\bibliography{references}



\end{document}